\documentstyle[12pt]{article}
\input epsf
\begin{document}\setlength{\unitlength}{1mm}

\def\question#1{{{\marginpar{\small \sc #1}}}}
\newcommand{\QCD}{{ \rm QCD}^{\prime}}
\newcommand{\MSSM}{{ \rm MSSM}^{\prime}}
\newcommand{\eq}{\begin{equation}}
\newcommand{\en}{\end{equation}}
\newcommand{\bino}{\tilde{b}}
\newcommand{\tsquark}{\tilde{t}}
\newcommand{\gluino}{\tilde{g}}
\newcommand{\photino}{\tilde{\gamma}}
\newcommand{\wino}{\tilde{w}}
\newcommand{\mtilde}{\tilde{m}}
\newcommand{\higgsino}{\tilde{h}}
\newcommand{\gsi}{\,\raisebox{-0.13cm}{$\stackrel{\textstyle>}
{\textstyle\sim}$}\,}
\newcommand{\lsi}{\,\raisebox{-0.13cm}{$\stackrel{\textstyle<}
{\textstyle\sim}$}\,}
\rightline{RU-96-26}
\rightline{hep-ph/9612346}
\rightline{December, 1996}
\baselineskip=18pt
\vskip 0.6in
\begin{center}
{ \LARGE SUSY and the Electroweak Phase Transition}\\
\vspace*{0.6in}
{\large Glennys R. Farrar and Marta Losada\footnote{On leave of absence
from Centro Internacional
de F\'{\i}sica and Universidad Antonio Nari\~{n}o, Santa Fe de
Bogot\'a, COLOMBIA.}} \\
\vspace{.1in}
{\it Department of Physics and Astronomy \\ Rutgers University,
Piscataway, NJ 08855, USA}\\
\vspace{.2in}
\vspace{.1in}
\end{center}
\vspace*{0.05in}
\vskip  0.2in  

Abstract: We analyze the effective 3 dimensional theory previously
constructed for the MSSM and
multi-Higgs models to determine
the regions of parameter space in which the electroweak phase transition
is sufficiently
strong for a $B+L$ asymmetry to survive in the low temperature phase. We
find
that the inclusion of all supersymmetric scalars and all 1-loop
corrections 
 has the effect of enhancing the strength
of the phase transition. Without a light stop or extension of the MSSM
 the phase transition is sufficiently first order only if
the lightest Higgs mass $M_{h}\lsi 70$ GeV and $tan\beta\lsi 1.75$.

\thispagestyle{empty}
\newpage
\addtocounter{page}{-1}
\newpage

 For electroweak baryogenesis to occur it is necessary that the
electroweak phase transition be
 sufficiently strongly
first order. Otherwise, sphaleron transitions after the phase
transition  wash out any baryon asymmetry which may have 
been produced at the electroweak scale \cite{kuzmin}. It is now known
that this requirement is not
satisfied in the minimal Standard Model \cite{KLRS, kajantie}. Thus, it
is of interest to investigate
extensions of the minimal Standard Model. Many authors have studied the
order of the
electroweak phase transition in the Minimal Supersymmetric Standard
Model (MSSM). 
Most of these studies rely on a one- and two-loop finite-temperature
effective potential 
analysis of the phase transition \cite{giudice, zwirner1, zwirner2,
carena1, espinosa1} in which stops were
expected to make the most significant contribution from supersymmetric
particles. 
The authors of these studies, in the limit of a large pseudoscalar Higgs
mass, $m_{A}\rightarrow \infty$, have identified a region of parameter
space
for which the transition is strong enough.  This corresponds to low
values of  $tan\beta$, and values of 
the soft supersymmetry breaking right stop mass, $m_{U_{3}}^{2}$, which
are small or even negative.  In reference \cite{zwirner2} the analysis
was extended for the full range of allowed values of $m_{A}$. It was  found
that larger values of $m_{A}$ are favored.

A different approach consists of separating
 the perturbative and non-perturbative aspects of the phase transition.
This is performed
 through the perturbative construction 
of effective three dimensional theories, and a subsequent lattice
 analysis of the reduced theory \cite{KLRS, kajantie, kks, farakos}. 
For the case in which the reduced theory
 contains a single
light Higgs field, characterized by a Higgs self-coupling,
$\bar{\lambda}_{3}$, and an effective  3D gauge 
coupling, $g_{3}$, the condition for a sufficiently strong first order
phase 
transition  becomes \cite{KLRS}

\begin{equation}
x_{c} = {\bar{\lambda}_{3}\over g_{3}^{2}} \lsi 0.04,
\label{xcrit}
\end{equation}
where the quantities $\bar{\lambda}_{3}$ and $g_{3}$ are functions of
the various parameters appearing in the 
original 4D theory.
  
An analysis of the parameter space for the reduced theory of the
Standard Model was  
performed in \cite{KLRS, kajantie}. It indicated  that for no value of
the Higgs mass is electroweak 
baryogenesis possible. This demonstrates that a purely perturbative
analysis is inadequate,
 since with that method the electroweak phase transition was
found to be sufficiently strong  for small values of the Higgs mass.
 In \cite{mlosada1} a 3D theory for the MSSM was constructed, including
Standard Model particles and
 additional
corrections arising from gauginos, higgsinos and all squarks and
sleptons. 
Here we use these results to  explore the MSSM parameter space
 in order to determine the regions for which electroweak baryogenesis
may occur\footnote{Throughout, we work in the $sin^2 \theta_W = 0$
approximation, which was found in ref. \cite{KLRS} to be adequate for
the MSM.}.
This relies on the relation between
the running parameters in the original 4D theory and physical
parameters; this is given 
 in  \cite{thesis}. 
A simplified version of the present analysis was performed by
others \cite{cline, laine}.  However only the contribution from gauge
bosons, higgses and third generation quarks and squarks to the 3D
reduction were included, and  one-loop corrections to 4D parameters
were not fully incorporated.  
In our work all one-loop corrections  as well as contributions from all
SUSY particles have been considered. This allows us to investigate the
effect of the full complement
of supersymmetric particles, 
in addition to third generation squarks, on
the strength of 
the phase transition. We find that these effects are important and
should not be neglected.
We also clarify the discrepancy between the results of references
\cite{cline} and \cite{laine}.
 In reference \cite{laine} the results
agreed basically with those found in the perturbative effective
potential analysis.
The most favorable region of parameter space was found to be $m_{h}\leq
m_{W}$ (low $tan\beta$), small stop mixing, 
$m_{U_{3}}\leq 50$ GeV and $m_{A} \geq 200$ GeV.
In addition to this region, reference \cite{cline}  found another region
of parameter space in which 
 arbitrary values of $tan\beta$
and a range of values for the pseudoscalar Higgs mass,  $40\leq m_{A}
\leq 80$ GeV, give a sufficiently
strong phase transition. We comment on the latter region below.

\hspace*{2em} The ratio $\bar{\lambda}_{3}/ g_{3}^{2}$ appearing in
equation
(\ref{xcrit}) depends on the parameters in the 4D theory 
($x_{c} = x(M_{A}, m_{o}, \mu, m_{{1\over 2}}, m_{\tilde{g}},
 A, tan\beta,T_{c})$), as well as  the Standard Model gauge couplings.
  $A$  is the scalar trilinear soft SUSY breaking
parameter, taken to be universal; $\mu$ is the supersymmetric higgsino 
mass parameter;
 $ m_{o}, m_{{1\over2}}, m_{\tilde{g}}$  denote the
common squark/slepton mass at the SUSY breaking scale \footnote{We
consider non-universal
squark masses where indicated.}, the  $SU(2)$ gaugino,  and  gluino
mass,  respectively.
 $M_{A}$ is the physical pole mass of the pseudoscalar Higgs, and
$tan\beta $ is the 
ratio of the vacuum expectation values of the Higgs fields in the
renormalized zero temperature
theory.
$x$ also depends indirectly on the scale $M_{SUSY}$, the scale at which
the SUSY boundary conditions on the quartic Higgs couplings  appearing
in the Higgs potential
are imposed \cite{hempfling} and at which
 mass parameters for squarks/sleptons are specified. 

The critical temperature, 
$T_{c}$, is defined to be the temperature at which there is  a direction
in field
space
at the origin of the Higgs potential for which the transition to the
minimum of the potential
in the broken phase  
can occur classically \footnote{We have checked that the difference in
the critical temperature from the diagonalization of
 the Higgs mass matrix,
 equation  (10) in \cite{mlosada1}  and from equation (7.9) in
\cite{laine} is extremely
 small ($\leq .1$ GeV) and for our purposes
negligible.}.
In 3D lattice calculations \cite{KLRS, farakos} the critical temperature
is taken to be
 the temperature of phase coexistence. These two values of temperature
are generally close but not identical. The actual  temperature at which
the phase transition occurs
lies between these two values. We will remark below on the circumstances
under which there can be a significant
difference arising from this distinction.

Throughout our analysis  we will concentrate on the regions of
parameter space which describe an effective theory in which there is a
single light scalar and thus the bound given by equation (\ref{xcrit})
is valid. However, we mention that another possibility is a scenario
in which two scalars, e.g. one Higgs and the right stop, are both
nearly massless at $T_{c}$ \cite{carena1}. A lattice calculation for
this extended 3D model is required before that scenario can be
investigated with the present approach.

\hspace*{2em} As is well known, the MSSM Higgs sector   can be
parametrized in terms of two quantities: 
$tan\beta$ and the pole mass of the pseudoscalar Higgs boson, $M_{A}$.
These are the most
important parameters in determining the strength of the phase
transition.
We consider $M_{A}$ values between 40-300 GeV in order to be compatible
with experimental limits and ensure the
 validity of
the high-temperature expansion.
In the figures we show results for $tan\beta$ between $1.25$ and $13.3$.
For larger and smaller values the
conclusions are essentially the same as for the extrema of the range
\footnote{In the dimensional reduction
procedure, the explicit dependence on 
all Yukawa couplings was kept. For the numerical
results presented here only the top Yukawa coupling is kept.
 We have explicitly checked that even
for large values of
$tan\beta$ the bottom Yukawa coupling can  be neglected.}.

In general, the masses of all particles are taken such that the high
temperature expansion
is valid\footnote{The adequate suppression of non-renormalizable
terms must also be verified.}.
The experimental constraints we impose on the masses are: for stop
masses
 $m_{\tilde{t}_{2}} \gsi 50$ GeV,  $m_{\tilde{t}_{1}} \gsi m_{t}$,
 for first and second generation squarks $m_{\tilde{q}_{i}} \gsi 200$
GeV, sleptons
 $m_{l} \gsi 50$ GeV, the gluino
mass either $\lsi 1$ GeV or  $\gsi 150$ GeV \cite{abe, farrar}. In
addition, the value of the
left soft supersymmetry breaking stop mass, $m_{Q_{3}}$, must be such that
the contribution from stops
and sbottoms to the $\rho$ parameter is not too large \cite{zwirner1}.

\begin{figure}
\vskip -100pt
\epsfxsize=4in
\epsfysize=6in
\epsffile{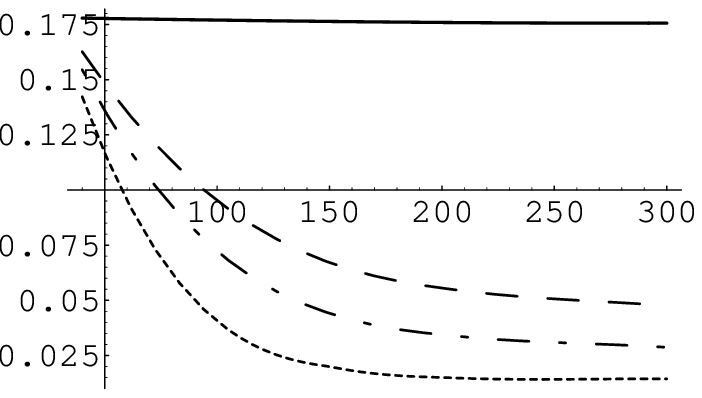}
\vskip -100pt
\caption{Plot of $x_{c}$ vs. $M_{A}$ for several different values of
$tan\beta$. The solid line
corresponds to $tan\beta= 13.3$, the dashed line to $tan\beta = 1.75$,
the dashed-dot line to
$tan\beta=1.5$, and the dotted line to $tan\beta = 1.25$.}
\label{xctanbeta}
\end{figure}

Figure \ref{xctanbeta} shows the value of $x_{c}$, for the case of no
squark mixing,
 as a function of the pseudoscalar Higgs mass for  values
of $tan \beta$ ranging from 1.25 to 13.3. We have fixed the other
parameters to be
$m_{o}=50$ GeV, $m_{{1\over 2}}= 50$ GeV, $m_{\tilde{g}}=
{\alpha_{s}\over \alpha_{W}}m_{{1\over 2}}$,
 $M_{weak}= m_{t}$,
$M_{SUSY} = 10^{12}$ GeV \footnote{In our approximation the masses of
sleptons and of the first and 
second generation squarks are fixed by $m_{o}$, $m_{\tilde{g}}$,
$m_{{1\over 2}}$,
through the renormalization group running; they
are constant as we vary $tan\beta$ and $M_{A}$. However, due to their
dependence on the renormalization group
running of the top Yukawa coupling, the left and right stop masses
change as we move
on the curves plotted in figure \ref{xctanbeta}.}.
For large values of $tan \beta$, there is no value of $M_{A}$ for which
$x_{c}$  fulfills 
the condition given by equation (\ref{xcrit}), and $x_{c}$ varies very
little as a function of
$M_{A}$. However, for low values of  $tan \beta$ and  large enough
values of $M_{A}$, $x_{c}$  can be small enough for the phase
transition to be sufficiently first order. 
The strong dependence of $x_{c}$ on the value of the pseudoscalar Higgs
mass, for 
low $tan\beta$, arises basically through the dependence
of the quantity $\bar{\lambda}_{3}$ in equation (\ref{xcrit}) 
 on the mixing angle $\theta$
 (see equation (18) in \cite{mlosada1}). It is easy
to see to lowest order the same dependence on $M_{A}$ arising in  finite
temperature effective potential analysis \cite{zwirner2}.
This qualitative  dependence of the strength of the phase transition on
$tan\beta$ and
$M_{A}$ was observed in \cite{zwirner1, zwirner2, espinosa1, laine}.
Varying the parameters $A$, $\mu$, $m_{\tilde{g}}$, $m_{{1\over 2}}$,
$m_{o}$,
$m_{t}$   either  increases $x_{c}$ slightly or has a negligible effect.

\begin{figure}
\vskip -100pt
\epsfxsize=4in
\epsfysize=6in
\epsffile{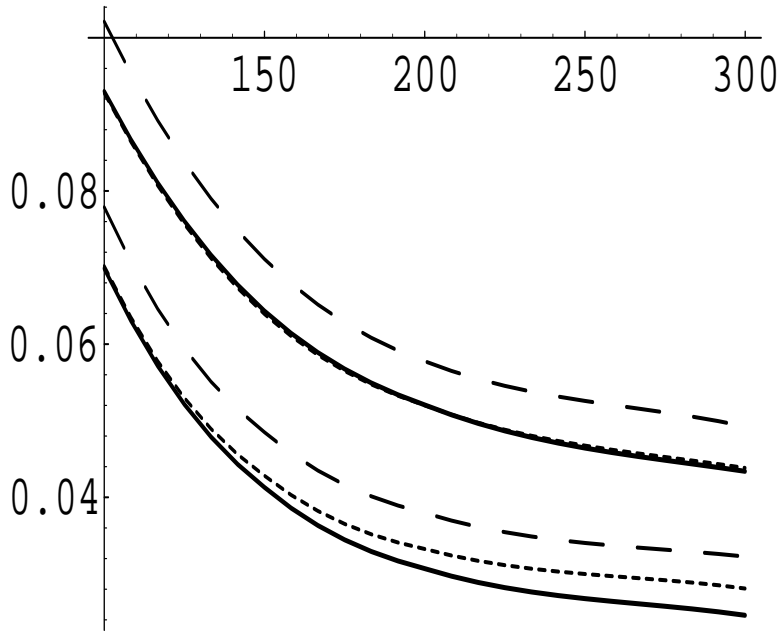}
\vskip -100pt
\caption{Plot of $x_{c}$ vs. $M_{A}$ for $tan\beta = 1.5$ and $1.75$
including all supersymmetric particles
(solid line), only third generation squarks (dashed lines) and all
squarks and sleptons but not 
gaugino corrections (dotted line).}
\label{comp3}
\end{figure}

We  have also compared the results of our general analysis to those
obtained with the simplifying
approximations used in \cite{cline, laine}, in which some supersymmetric
particles are neglected.
 In all cases we have kept the full one-loop
corrections to the 3D couplings, in contrast to \cite{cline, laine}.
Figure \ref{comp3}
 shows the variation of $x_{c}$ with $M_{A}$ for three
different cases and for two values of $tan\beta$. The solid line
 corresponds to our general analysis, as given above. The dashed line is
the result when only  third generation squarks are included; the gluino
and electroweak gaugino thermal
screening contribution to the 3D masses of the squarks
is  excluded. The dotted line corresponds to the case in which we 
include  all  squarks and sleptons, but ignore
 all gaugino contributions to the three dimensional theory. Although the
effect of the right stop
on the strength of the phase transition is greater than that of any
other sfermion, the ensemble
of sfermions neglected in \cite{cline, laine} significantly strengthens
the phase transition. As expected
from the work of \cite{zwirner2, carena1, laine}, we find that the
reduction of
the right stop soft supersymmetric breaking mass  decreases 
$x_{c}$.   However, the decrease is less important than it appears when
only the contribution from third generation
squarks \cite{laine} is included. 
We have also compared the two cases in which only third generation
squarks were included with and without
thermal screening arising from the gluino and gaugino. The differences
in the values of $x_{c}$ for this case
are negligible.

For given squark masses at the SUSY scale, the running masses at low
energies are reduced as the gluino mass decreases.  As a result of
these lower masses the sfermions' favorable impact on the phase
transition is increased.  Thus light gluinos can be helpful in
providing a sufficiently strong ew phase transition, and the low
values of $tan \beta$ required for the phase transition lead to
chargino masses in an acceptable range in the light gaugino
scenario \cite{f_chargino}.  

An important point, which has been overlooked in the previous
literature, is that for some regions of parameter space it may be
incorrect to conclude from the above analysis that the phase
transition is not sufficiently strongly first order. As noted
previously the actual transition temperature is somewhat higher  than
$T_{c}$ as defined above.  For some values of $M_A$ and $tan \beta$,
$x_{c}$ depends strongly on the temperature.  This happens when the
mixing angle $\theta$, which diagonalizes the 3D Higgs mass matrix at
finite temperature, varies rapidly for temperatures near the critical 
temperature.  A rapid variation of the mixing angle occurs when the
temperature is such that the diagonal elements  of the 3D Higgs mass
matrix become nearly degenerate. If the critical temperature for the
phase transition is  close to the value of the temperature where this
rapid variation occurs,  then the value of $\theta$ and consequently
of $x_{c}$ are very sensitive to the transition temperature. Note that
in general our procedure of integrating out the heavier Higgs is not
compromised when this phenomenon occurs because the eigenvalues remain
well-separated. Rather, our inability to analyze this region of
parameter space arises from our inability (with present techniques)
to obtain a sufficiently accurate determination of the phase
transition temperature. 

\begin{figure}
\vskip -100pt
\epsfxsize=4in
\epsfysize=6in
\epsffile{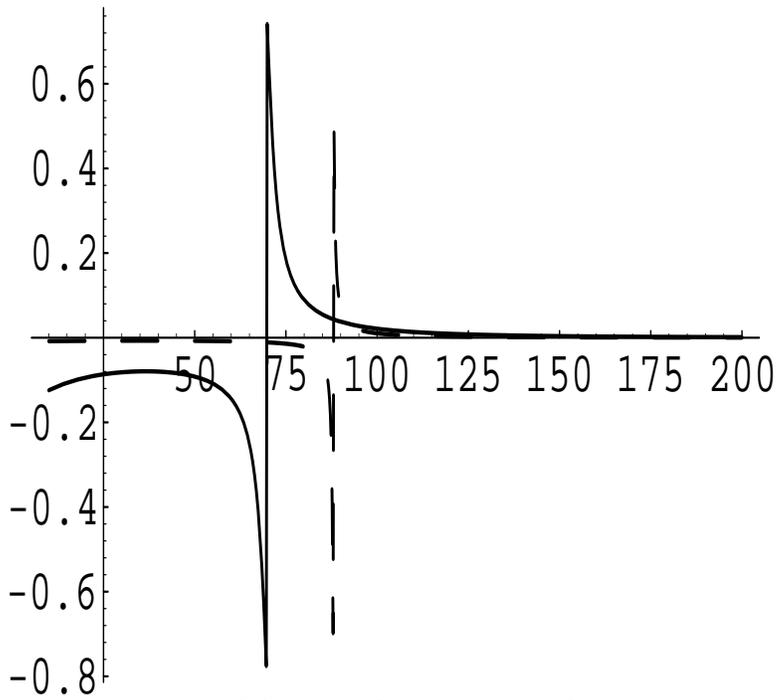}
\vskip -100pt
\caption{Plot of $\theta$ vs. $T$ for  $M_{A} = 40$ GeV. The solid line
corresponds to
 $tan\beta = 1.25$, $T_{c}= 63.3$ GeV,  the dashed line is for $tan\beta
= 13.3$, $T_{c}= 75.6$ GeV.}
\label{thetaT40}
\end{figure}

\begin{figure}
\vskip -100pt
\epsfxsize=4in
\epsfysize=6in
\epsffile{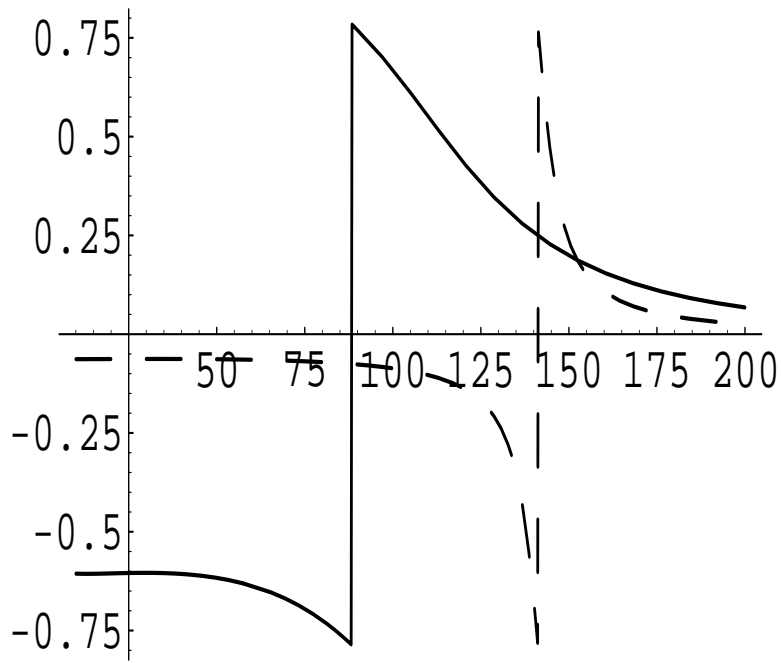}
\vskip -100pt 
\caption{Plot of $\theta$ vs.$T$ for $M_{A} = 300$ GeV, $T_{c} =
60,75.6$ GeV. The solid line
corresponds to $tan\beta = 1.25$, the dashed line to $tan\beta = 13.3$. 
}
\label{thetaT300}
\end{figure}

\begin{figure}
\vskip -100pt
\epsfxsize=4in
\epsfysize=6in
\epsffile{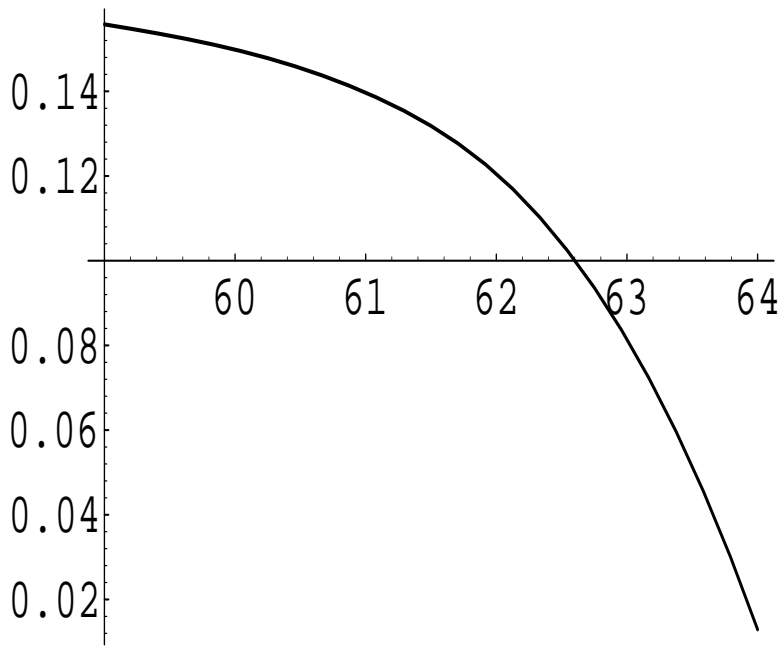}
\vskip -100pt
\caption{Plot of $x_{c}$ vs.$T$ for $tan\beta = 1.25$, $M_{A} = 40$ GeV,
$T_{c}= 60.8$ GeV. }
\label{xcT40}
\end{figure}

The value of the temperature at which this rapid variation occurs
depends strongly 
on the value of the pseudoscalar Higgs mass. As $M_{A}$ increases this
 temperature also increases; for $M_{A}\gsi 100$ GeV it is
well-separated from the phase transition temperature.
 Moreover, the extent of the variation of $x_{c}$ is less for larger
$tan\beta$.
 Figure  \ref{thetaT40} shows  the dependence of
$\theta$ on the temperature for $M_{A}=40$ GeV;
 the solid line corresponds to  $tan\beta= 1.25$
and the dashed line to $tan\beta= 13.3$, keeping all other parameters
fixed.
 Figure \ref{thetaT300}
is the same  for $M_{A}=300$ GeV. Figure \ref{xcT40} shows $x_{c}$ as a
function of
 temperature close
 to $T_{c}$, for $M_{A}=40$ GeV. 
A $5$ GeV variation  in the temperature around $T_{c}$ induces, for
$M_{A}=40$ GeV,
a change in the value
 of $x_{c}$, $\Delta x_{c}\sim .13$, while 
for $M_{A}= 300$ GeV,  $\Delta x_{c} \sim 0.005$.
Thus the possibility of large uncertainty in the mixing angle is
relevant only for low values of $tan\beta$ and  $M_{A}$. Since
$M_{A}\lsi 100$ GeV is already ruled out experimentally \cite{aleph}
for the MSSM in the $tan\beta$ region of interest, this possibility
cannot enlarge the viable region of parameter space in the MSSM and
we do not pursue it further. However we note that this phenomenon may
play a role in the discrepancy between the conclusions of \cite{cline,
laine} for $M_{A}\lsi 100$ GeV. 

\begin{figure}
\vskip -100pt
\epsfxsize=4in
\epsfysize=6in
\epsffile{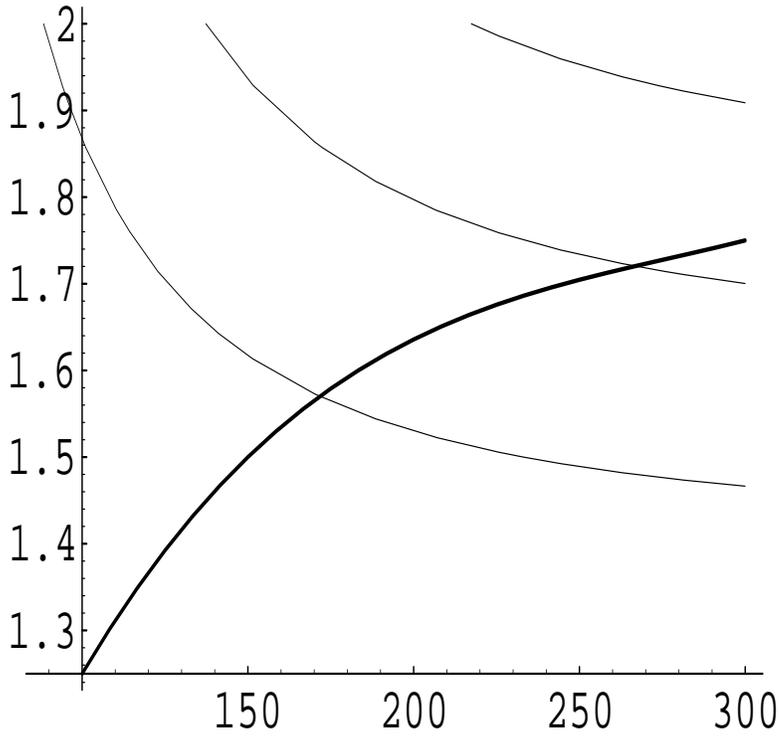}
\vskip -80pt
\caption{Contours of constant $M_{h} = 70,60,45$  from top to bottom in
the $tan\beta$ vs $M_{A}$ plane. 
The thick line gives the
constraint for a sufficiently strong first order phase transition,
$x_{c}=0.04$.}
\label{contours}
\end{figure}

We now turn to implications of these constraints for the
mass of the lightest Higgs.  The light curves in figure \ref{contours}
indicate contours of constant $M_{h}$ in the $tan\beta - M_{A}$ plane,
using the results given in \cite{thesis} 
to relate the running parameters of the MSSM analysis to the physical
parameters and taking $m_{\tilde{t}_{2}} \sim 180$ GeV,  
$m_{\tilde{t}_{1}} \sim 320$ GeV\footnote{This choice of parameters
maximizes the region in
$tan\beta - M_{A}$ giving $x_c \le 0.04$.}.  Due to approximations in
the RG analysis and our ignorance with respect to the mass of the
stop, one should attach a few-GeV uncertainty to these curves.  The
region of $tan\beta - M_{A}$ which preserves a baryon asymmetry
generated at the weak scale is below and to the right of the thick
line, which corresponds to $x_{c} =0.04$.  We conclude that unless
$M_{h}\lsi 70$, electroweak baryogenesis is not viable in the MSSM. 

Given that experimental Higgs limits already nearly exclude such a
light Higgs, we briefly mention three alternate extensions of the
SM which might allow electroweak baryogenesis.  In order to increase
the strength of the phase transition one must either increase the
$\phi^3 T$ term in $V_{eff}$, or decrease the coefficient of the
$\phi^4$ term, or both.  The first possibility is employed in theories
which are fine-tuned such that one or more scalars {\it in addition
to} the usual Higgs scalar are nearly massless at the phase transition
temperature.  Terms in $V_{eff}$ cubic in $\phi$ arise from 
bosonic mass-squareds which are proportional to $\phi^2$.  Such
contributions are enhanced if there is a precise cancelation of the
thermal contributions to the mass, e.g., from a negative mass-squared at
the SUSY scale \cite{carena1}.  Other scalars such as additional Higgs
or sneutrinos could in principle be employed to serve a similar
purpose, although the stop is particularly natural and has 3 color
degrees of freedom as well.  Non-perturbative effects in such
scenarios cannot be analyzed without further lattice calculations.
For the light stop scenario, the 3D theory which must be analyzed on
the lattice is considerably more complicated than the one relevant to
the SM \cite{KLRS} and the theories analysed here.  The relevant 3D
lattice calculation for the light stop scenario must include SU(3) as
well as SU(2) gauge interactions, and one of the scalars couples to both
SU(2) and SU(3) gauge bosons, as well as to the other 
scalar.  See ref. \cite{laine, mlosada1} for the full 3D reduction with two
light scalars. 

The other strategy to enhance the phase transition is to decrease the
coefficient of the $\phi^4$ term, i.e., $\bar{\lambda}_3$ in the
3D theory. This may be possible either in non-SUSY multi-Higgs doublet
models or in the NMSSM (minimal SUSY augmented with a gauge singlet
Higgs).  Non-SUSY multi-Higgs theories have some advantage in this
regard, since the Higgs potential of the 4D theory is not fixed by
gauge couplings at the SUSY scale.  It is of course also constrained
by non-observation of a Higgs particle and the requirement that the
broken-symmetry vacuum is the minimum of the T=0 theory.  A
disadvantage of non-SUSY theories is the absence of sfermions, which
enhance the strength of the transition, as we saw above.  Finally, the
NMSSM has sfermions and also more freedom in the Higgs sector, so the
lightest Higgs in the 4D theory may be acceptably heavy even with a
small $tan \beta$, without requiring a large $M_A$ \cite{nmssm1,
nmssm2}.  For this theory in particular the $\theta$ dependence noted
above may prove important in the analysis. 

In conclusion, we have employed existing lattice calculations to
analyze the electroweak phase transition in the MSSM, including
non-perturbative as well as perturbative thermal effects.  We include all 
1-loop corrections and integrate out all
gauginos and sfermions.  Although we find qualitative agreement with
the results of refs. \cite{cline,laine}, we find that inclusion of all
sfermions and D-term couplings is important quantitatively, because of
their large multiplicity.  This
enhances the strength of the phase 
transition in the relevant regime of parameters.  We find
that the MSSM provides sufficient suppression of sphaleron transitions
in the broken phase only for values of $tan\beta \lsi 1.75$, unless
the right stop soft-SUSY breaking mass is less than $50$ GeV.  In the
latter case coefficients of non-renormalizable terms become large,
signaling the onset of the breakdown of our analysis when there are two
light scalars at the phase transition.  In this regime our method does
not apply, although purely perturbative analysis leads one to expect
that the strength of the phase transition is enhanced
\cite{zwirner2,carena1}.  The strength of the 
phase transition increases as the mass of the pseudoscalar Higgs
increases.  $M_{A}$ can be as low as $100$ GeV, for $tan\beta \sim
1.25$, and still give $x_{c}\lsi 0.04$, even assuming universal soft 
supersymmetry breaking masses at the SUSY breaking scale (i.e.,
without the light stop scenario).  The region of parameter space
potentially supporting electroweak baryogenesis requires the lightest
physical Higgs mass $M_{h}$ to be $ \lsi 70$ GeV.  This is very close
to being experimentally excluded.  We commented on possible
alternatives to the MSSM in which electroweak baryogenesis could be
compatible with experimental constraints on the Higgs mass.

\vspace{.2in}
\noindent
{\bf Acknowledgments:} 
 We thank M. Shaposhnikov  for many useful discussions. 
We also thank M. Laine for comments on the manuscript and P. Janot
for e-mail
correspondence.
 Research supported in part by NSF-PHY-94-23002 and Colciencias,
Colombia.


\newpage

\vspace{.2in}

\end{document}